\pdfoutput=1
\documentclass[a4paper,american,floatfix,pdftex,superscriptaddress,twoside,%
aip,jcp,%
citeautoscript,
reprint, final
]{revtex4-2}%
\usepackage[T1]{fontenc}
\usepackage{graphicx}%
\usepackage[utf8]{inputenc}
\usepackage{hyperref, hypernat}
\usepackage[todos]{CMImacros}
\usepackage[boxed]{algorithm2e} 
\usepackage{amsfonts,amsmath,amssymb}%

\newcommand{\cfeldesy}{\affiliation{Center for
Free-Electron Laser Science CFEL, Deutsches
      Elektronen-Synchrotron DESY, Notkestraße 85, 22607 Hamburg, Germany}}%
\newcommand{\uhhchem}{\affiliation{Department of Chemistry, Universität Hamburg,
      Martin-Luther-King-Platz 6, 20146 Hamburg, Germany}}%
\newcommand{\uhhcui}{\affiliation{Center for Ultrafast Imaging, Universität Hamburg, Luruper
      Chaussee 149, 22761 Hamburg, Germany}}%
\newcommand{\uhhphys}{\affiliation{Department of Physics, Universität Hamburg, Luruper Chaussee 149,
      22761 Hamburg, Germany}}%
\newcommand{\uhhmaths}{\affiliation{Department of Mathematics, Universität Hamburg, Bundesstraße 55,
      20146, Hamburg, Germany}}%
\newcommand{\ayemail}{\email[Email: ]{andrey.yachmenev@cfel.de}}%
\newcommand{\cmiweb}{\homepage[URL:~]{https://www.controlled-molecule-imaging.org}}%

\SetAlCapSkip{\abovecaptionskip}
\begin{document}
\title{Active learning of potential-energy surfaces of weakly-bound complexes with regression-tree
   ensembles}%
\author{Yahya Saleh}\cfeldesy\uhhmaths %
\author{Vishnu Sanjay}\cfeldesy\uhhmaths\uhhcui%
\author{Armin Iske}\uhhmaths%
\author{Andrey Yachmenev}\ayemail\cmiweb\cfeldesy\uhhcui%
\author{Jochen Küpper}\cfeldesy\uhhcui\uhhchem\uhhphys%
\date{\today}
\begin{abstract}
   Several pool-based active learning algorithms (AL) were employed to model potential energy
   surfaces (PESs) with a minimum number of electronic structure calculations. Theoretical and
   empirical results suggest that superior strategies can be obtained by sampling molecular
   structures corresponding to large uncertainties in their predictions while at the same time not
   deviating much from the true distribution of the data. To model PESs in an AL framework we
   propose to use a regression version of stochastic query by forest, a hybrid method that samples
   points corresponding to large uncertainties while avoiding collecting too many points from sparse
   regions of space. The algorithm is implemented with decision trees that come with relatively
   small computational costs. We empirically show that this algorithm requires around half the data
   to converge to the same accuracy in comparison to the uncertainty-based query-by-committee
   algorithm. Moreover, the algorithm is fully automatic and does not require any prior knowledge of
   the PES. Simulations on a 6D PES of \pyrrolew show that $\mathord{<}15\,000$ configurations are
   enough to build a PES with a generalization error of 16~\invcm, whereas the final model with
   around 50\,000 configurations has a generalization error of 11~\invcm.
\end{abstract}
\maketitle

\section{Introduction}
The ability of molecules to form weakly-bound complexes, \emph{via} van der Waals or hydrogen bonds,
is fundamental to many physical, chemical, and almost all biological processes, ranging from the
phenomenon of resonances in reactive scattering~\cite{Liu:ACP149:1} and roaming in chemical
reactions~\cite{Suits:ARPC71:77} to biochemical processes in aqueous
solution~\cite{Colombo:Science256:655}. Small microsolvated complexes of aromatic molecules or
chromophores are important model systems for studying the interactions established by individual
water molecules at specific binding sites~\cite{Zwier:ARPC47:205}. They offer an appealing way for
the quantitative investigations of the effects of hydration on the photoexcitation and ionization
dynamics and the mechanisms of chemical reactions in general~\cite{Korter:JPCA102:7211,
Hertel:RPP69:1897, Kang:IRPC24:1,
   Samanta:CR116:4913, Johny:protection:inprep, Onvlee:indole:inprep}.

First principles calculations of potential energy surfaces (PESs) for weakly-bound complexes are
challenging and computationally expensive~\cite{Akin-Ojo:JCP123:134311, Metz:PCCP21:13504,
   Ma:JCTC15:1044, Metz:JCTC16:2317}. The presence of small energy differences in weakly-bound PES
means that generally high levels of theory need to be employed to produce correct asymptotic
behavior~\cite{Vogels:NatChem10:435}. Furthermore, the landscape of these PESs is complex because of
the loosely bound character of intermolecular interactions. Thus, a larger number of grid points is
generally required to sample the complete configuration space. Moreover, due to the importance of
dynamical electron correlation (dispersion) and its slow basis-set
convergence~\cite{Brauer:PCCP18:20905}, calculations for the noncovalent long-range parts of the PES
are generally more costly than the ones at short-range.

The standard approach for building \emph{ab initio}-based molecular PESs employs a regression
procedure: one generates a grid of distinct molecular geometries at which the potential energy is
computed, then a model is fitted to these data by minimizing a loss function, \eg, root-mean-square
(RMS) error, under a regularization constraint~\cite{Behler:JCP145:170901, Hansen:JCTC9:3404,
   Manzhos:CR:inprep}. In recent years, many machine learning (ML)
models~\cite{Blum:FounDataSc:2020, Lotz:MathML:2020, Yao:MathDS:2019, Shalev:ML:2014,
   Goodfellow:DL:2016} have been used to fit molecular PESs. The most extensively used models
include permutationally invariant polynomials~\cite{Wang:PTRSA375:20160194, Braams:IRPC28:577,
   Xie:JCTC6:26, Qu:ARPC69:151, Conte:JCTC16:3264}, neural networks
(NNs)~\cite{Morawietz:JCP136:064103, Behler:PRL98:146401, Unke:JCTC15:3678, Manzhos:IJQC115:1012,
   Behler:IJQC115:1032, Jiang:IRPC35:479, Schran:JCTC16:88}, Gaussian processes
(GPs)~\cite{Kamath:JCP148:241702, Bartok:PRL104:136403, Qu:JCTC14:3381, Rasmussen:GaussProc:2006,
   Sugisawa:JCP153:114101, Dai:JCTC16:1386, Vargas:NJP21:022001}, and other kernel
methods~\cite{Unke:JCIM57:1923, Dral:JCP146:244108, Koner:JCTC16:5474, Iske:ApproxTh:2018}.

The quality of PESs strongly depend on the chosen set of molecular geometries sampled. This choice
is often based on humans' intervention: the regions that will likely be intensively accessed by the
dynamics simulations are sampled more densely. If one's assumptions about the distribution of
geometries accessed in dynamics' simulations are mostly correct, the question is what minimal amount
of data points is needed to reproduce the PES to a desired accuracy.

Trying to minimize the number of electronic structure calculations and reducing the amount of
humans' intervention in an active learning (AL) paradigm~\cite{Settles:ActiveLearning:2009} became
increasingly popular during the last few years~\cite{Peterson:PCCP19:10978, Lin:JCP152:154104,
   Zhang:PRM3:023804, Uteva:JCP149:174114, Loeffler:JPCC124:4907, Zhai:JCP152:144103,
   Vandermause:CM6:1, Podryabinkin:CMS140:171, Gastegger:CS8:6924, Smith:JCP148:241733,
   Sivaraman:CM6:1, Guan:MP116:823, Lin:JCTC17:2691}. AL
   encompasses a variety of iterative algorithms aimed at
minimizing the cost of training data acquisition. In pool-based AL, the expert defines a pool of
unlabeled data, \eg, molecular geometries, and a policy algorithm queries the energies of geometries
that would most significantly minimize the generalization error of the PES once labeled and included
in the training data~\cite{Settles:ActiveLearning:2009}.
The vast majority of AL
applications to PESs used uncertainty-based sampling~\cite{Peterson:PCCP19:10978, Zhang:PRM3:023804,
   Uteva:JCP149:174114, Loeffler:JPCC124:4907, Zhai:JCP152:144103, Vandermause:CM6:1,
   Podryabinkin:CMS140:171, Gastegger:CS8:6924, Smith:JCP148:241733}, which queries points
corresponding to the largest uncertainties in their predicted
energies. For probabilistic models
like GPs, the uncertainties can be directly
calculated~\cite{Zhai:JCP152:144103, Vandermause:CM6:1,
Uteva:JCP149:174114, Guan:MP116:823}.
For the ML models that do not offer a direct way to compute uncertainties, these can be inferred by
training an ensemble of models on the currently labeled training set and selecting the points about
which the models disagree the most. This algorithm is called query-by-committee
(QBC)~\cite{Seung:CLT5:287} and it was used for uncertainty-based sampling of
PESs~\cite{Zhang:PRM3:023804, Gastegger:CS8:6924, Lin:JCP152:154104, Smith:JCP148:241733,
   Peterson:PCCP19:10978}.

Uncertainty-based sampling is known to be prone to querying outliers~\cite{Kee:INSC454:401,
   Settles:ActiveLearning:2009}, because it does not take into account the statistical distribution
of data in the pool and hence partly ignores the user’s \emph{a priori} knowledge embedded in the
pool. Upper bounds on the generalization error in an AL setting suggest that more optimal AL
strategies can be built by sampling points corresponding to large
uncertainties while not deviating
much from the true distribution of
data~\cite{Wang:ACMTKDD9:23, Shui:AISTATS2020:1308}. In
applications to PESs, uncertainty-based algorithms tend
to select many molecular geometries from regions sparsely sampled by subsequent dynamics
calculations. When all sparse regions of the pool can be clearly identified, \eg, as points with
high energy, or if prior knowledge about the minima and saddle points exists, this problem can be
solved by introducing a weighting function
\cite{Guan:MP116:823, Lin:JCP152:154104}. In a more general
setting, one can combine the
uncertainty-based query algorithm with a molecular-dynamics sampler starting from various known
critical points of the PES \cite{Zhang:PRM3:023804, Gastegger:CS8:6924, Wang:PTRSA375:20160194}.

Another concern in AL is the execution speed and the scaling with
the amount of data, as AL is an iterative procedure that includes
training and predicting energies of many geometries in
each iteration.

Here, we propose a regression-version of the stochastic query by forest (SQBF)~\cite{Borisov:AL}
selection algorithm and use it to generate a minimal grid for the intermolecular PES of weakly-bound
\pyrrolew complex~\cite{Johny:CPL721:149, Johny:protection:inprep}. The algorithm selects molecular
geometries with large uncertainties in corresponding energy predictions and at the same time keeps
the distribution of selected geometries similar to that in the pool. The proposed algorithm does not
suffer from the shortcomings of uncertainty-based methods and leads to better results. The
uncertainty is estimated using an ensemble of regression trees that constitute a random forest
regressor (RFR), while the data statistics is partially preserved using a random sampling (RS)
procedure. Owing to the greedy optimization algorithm employed in regression trees, this method is
computationally fast and scales well with the size of data. In general, the RFR model is not a
continuous function and hence not well suited for modeling of PESs. However, due to its high
computational efficiency, the RFR is an attractive method for sampling purposes in AL, which can be
easily combined with other ML models, \eg, NNs, to produce the final model. Our empirical results
show that the SQBF selection scheme combined with a NN model for the fitting of PES require only
30~\% of the initial pool data (around 50\,000 grid points) to achieve the convergence with a
generalization error of about 16~\invcm. In comparison to popular QBC and RS schemes, the SQBF
queries about two and three times less data, respectively, to converge to the same accuracy. The
SQBF algorithm is general, not restricted to weakly-bound complexes, and applicable to larger
molecular systems.

\section{Methods}
\label{sec:methods}
\subsection{A formal look into active learning}
\label{ssec:AL_theory}
We denote the joint
distribution of the problem as $\cal{P}$, where $\mathcal{P}(x,E)=p(x)p(E|X)$. In
standard supervised ML, one has a dataset of points that are independently sampled and identically
distributed according to $\mathcal{P}$~\cite{Shalev:ML:2014}. In an AL paradigm, the dataset is
selectively sampled from another data distribution $\mathcal{Q}(x,E)$ that has the same conditional
distribution of $\mathcal{P}$, \ie, $\mathcal{Q}(x,E) = q(x) p(E|X)$~\cite{Shui:AISTATS2020:1308,
   Wang:ACMTKDD9:23}. We define $P=\{X_1 \dots X_l\}$ to be the set of all distinct molecular
geometries in the pool and assume that $P \sim p(x)$. We define
$S^{(0)}=\{(X_1, E_1)\dots (X_m, E_m) \}$ to be the initial labeled subset of $P$, where $m\ll{}l$.
Molecular geometries in the initially labeled set are denoted by $S^{(0)}_X := \{X_1 \dots X_m\}$,
so that $P^U = P \setminus S_X ^{(0)}$ determines unlabeled geometries in the pool. The steps of a
pool-based AL algorithm in a batch-mode~\cite{Settles:ActiveLearning:2009} are summarized
in \autoref{Alg:AL}.
\begin{algorithm}
   \SetAlgoLined%
   Fix batch size $B, t=1$ \;
   \textbf{Input:} Pool of unlabeled data $P^U$, an initial labeled data $S^{(0)}$ \\
   \While{performance is unsatisfactory}{
      a) Select B elements from $P^U$ to obtain $S^{(t)}_X$ \\
      b) Label this set to obtain $S^{(t)}$  \\
      c) Set $P^U = P^U \setminus S^{(t)}_X$, $S^{(t)} = S^{(t-1)} \cup S^{(t)}$ \\
      d) $t=t+1$}%
   \caption{Basic steps of a generic pool-based active learning strategy. In each active
   learning iteration, $B$ data points are chosen from the pool, labeled, and added to the training
   data.}
   \label{Alg:AL}
\end{algorithm}
To decide whether or not the performance is satisfactory, one needs to choose a stopping criterion.
We used the root mean square (RMS) error on a test dataset. We refer to an algorithm that performs
step $(a)$ as a policy algorithm.

A simple example of a policy algorithm is
uniform
RS
from the pool. Such an algorithm
is representativeness based, as it samples more
data from
dense regions. In the present case,
RS queries more molecular geometries with
energies close
to the dissociation limit,
which is due to the distribution of data in the pool.

An example of an uncertainty-based algorithm is QBC. In this algorithm one trains an ensemble of
efficient and diverse learners on the currently labeled dataset. To decide whether the inclusion of
a certain point would improve the performance of the regressor, these learners are asked to predict
the energies of this point. If they produce very different values, we conclude that adding this
point to the training set will improve the performance and query the user about the corresponding
energy. The algorithm is formalized in \autoref{alg:QBC}.
\begin{algorithm}[t]
   \SetAlgoLined%
   Fix the number of learners $n$ in the ensemble. \\
   \KwResult{$B$ elements from $P^U$}
   \textbf{Input:} $P^U$, $S^{(t)}, B$ \\
   a) Train an ensemble $\{T_i\}_{i=1}^n$ of models on data $S^{(t)}$. \\
   b) Compute predictions $\hat{E_i} = T_i (X) \; \forall X \in P^U, \; \forall i$\\
   c) Compute the community disagreement $q(X) = \text{std } (\hat{E_i}) \; \forall X \in P^U$ \\
   d) Take $B$ elements from the unlabeled data that have the highest $q(X)$ \\
   \caption{Basic steps of a query-by-committee algorithm for regression problems. $n$ learners are
      trained on the labeled datasets and asked to make predictions on the whole unlabeled dataset.
      The $B$ elements chosen to be labeled are those that maximize the standard deviation of the
      prediction among 5 learners.}
   \label{alg:QBC}
\end{algorithm}
We note that diversity of learners is crucial in this algorithm. If the learners are not diverse,
their predictions for a certain geometry would be almost the same and hence one would not be able to
infer the uncertainty. Practically, the diversity of learners is introduced through random
perturbations to the learning process.

Uncertainty-based algorithms do not take the statistical properties of data into account when
querying points, which is a serious drawback when the distribution of molecular geometries in the
grid is nonuniform. We empirically show that QBC tends to sample points from sparse regions to the
detriment of high-density parts. To correct this behavior, one can construct a probability density
function from the QBC-estimated uncertainties. Then, querying grid points is performed through
random sampling according to this density function. \autoref{alg:SQ} formalizes this idea.
\begin{algorithm}[t]
   \SetAlgoLined%
   Fix the number of learners $n$ in the ensemble. \\
   \KwResult{$B$ elements from $P^U$}
   \textbf{Input:} $P^U$, $S^{(t)},B$ \\
   a) Train an ensemble $\{T_i\}_{i=1}^n$ of $n$ models on data $S^{(t)}$.  \\
   b) Compute predictions  $\hat{E_i} = T_i (X) \; \forall X \in P^U, \; \forall i$ \\
   c) Compute the community disagreement $q(X) = \text{std } (\hat{E_i}) \; \forall X \in P^U$ \\
   d) Compute the weights:
   $L(X) = \frac{q(X) - q_\text{min}}{q_\text{max}-q_\text{min}} \text{, and the
      sampling probability } p(X) =
      \frac{L(X)}{\Sigma
      _ X
      L(X)}$ where $q_\text{min} =
      \underset{X}{\text{min
      }}
      q(X)$ and $q_\text{max} =
      \underset{X}{\text{max
      }}
      q(X)$\\
   e) sample $B$ elements from the unlabeled data with probabilities $p(X)$ \\
   \caption{Stochastic query-by-committee algorithm: Data to query are chosen by sampling according
      to a probability distribution that gives more weights to data points whose predictions are
      uncertain. Uncertainty is inferred by a standard query-by-committee algorithm.}
    \label{alg:SQ}
\end{algorithm}
In contrast to QBC, points with small uncertainties, such as for
example points close to the
asymptotic limit, may still be queried if they fall in high-density regions. In other words,
\autoref{alg:SQ} respects the statistical information in the pool that is defined \emph{a priori} by
the expert. Note that accounting for the statistical information in the pool using QBC can also be
performed by considering only a few unlabeled data points sampled independently from the input
distribution as candidates to query~\cite{Freund:ANIPS1993:483}, which is very similar in spirit to
\autoref{alg:SQ}. However, we empirically observed \autoref{alg:SQ} to work better than this
approach.

The balance between sampling points from the sparse and high-density regions is controlled by the
function $L$, which is linear with respect to the community
disagreement. The probability of
a point being sampled decreases linearly with the decrease of the point's uncertainty. One can have
more freedom on this balance by considering powers of this
function, \ie, $L^{\alpha}$,
where $\alpha\in\mathbb{R} ^+$. For $\alpha\in(0, 1)$, the algorithm will sample more points with
low uncertainty and conversely less for $\alpha\in(1,\infty)$. We
performed a heuristic study
of the effect of different powers $\alpha$. At each AL iteration, we ran SQBF algorithm for
different values of $\alpha\in\{0.5, 0.75, 1, 1.25, 1.75\}$. We picked the $\alpha$ that led to the
largest improvement in generalization error and collected the corresponding queried points. We
proceeded to active learning using this batch as part of the pool. The whole procedure was repeated
at every AL iteration. We found only minor improvements of the accuracy when using multiple,
optimized, values of $\alpha$. We explored a few other heuristics of similar nature, but none of
them yielded significantly better results. Hence, throughout the paper we report results obtained
with a single value of $\alpha=1$. In \autoref{alg:QBC} and \autoref{alg:SQ} it remains to specify
the ensemble of learners and to elaborate on how to diversify them. While any ensemble of ML models
can be used, we propose to use the trees of an RFR as members of this ensemble. In the next section
we argue that choosing regression trees for inferring uncertainty is advantageous because of
relatively low-training complexity and a straightforward diversification-ability. The RFR combined
with \autoref{alg:SQ} gives rise to a regression version of the stochastic query by forest algorithm
(SQBF)~\cite{Borisov:AL}, employed in this study.

\subsection{Random Forest Regressor}
\label{ssec:RFR}
Regression trees are a non-parametric way of solving a regression problem. They are based on the
intuition that the output value can be inferred by partitioning the input space. In particular, for
solving a regression problem with data pairs $\{(X_i, E_i)\}_{i=1}^{l}$, a \emph{Tree-Regressor}
aims at finding $J$ distinct and non-overlapping regions $R_1\ldots{}R_J$ in the feature space that
minimize
\begin{equation*}
   \sum_{j=1}^{J} \sum_{i; \; X_i \in R_j} (E_i -  \bar{E}_{R_j})^2
\end{equation*}
where $\bar{E}_{R_j} = \sum_{i; \; X_i \in R_j} E_i/m_j$ is the average energy of $m_j$ molecular
geometries in region $R_j$~\cite{James:IntroSLearning:2013}. This problem is NP-complete
\cite{Laurent:IPL5:15}. Therefore, only near-optimal solutions are considered by restricting
ourselves to hyper-rectangular regions and using recursive binary splitting, a greedy algorithm to
obtain a near-optimal segmentation. A prediction for a new molecular geometry is done by assigning
the input geometry to one of the regions. The prediction for this geometry is then the average
energy of all geometries in the training dataset that fall in this region. A major drawback of tree
regressors is their large variance~\cite{James:IntroSLearning:2013}. A powerful
approach to mitigate this problem is to consider an ensemble of trees. The key idea is that
averaging a set of independent random variables, which have comparable variances, reduces their
overall variance~\cite{James:IntroSLearning:2013}. In an ensemble method, a random perturbation is
introduced to the learning process in order to produce several different learners from the same
training set. Thus, taking the average of the predictions of the ensemble would result in a
reduction of variance. Such a random perturbation can be introduced by bootstraping, which gives
rise to bootstrap aggregation (bagging) methods. Here, $B$ different bootstrapped datasets of size
$m_b$ are generated. A tree is built on each model. For a new data point, a prediction is made by
taking the average of the predictions of all trees:
${T}_{\text{bag}} (x) = \frac{1}{B} \sum_{i=1}^{B} {T}_{i} (x)$.

A further random perturbation in tree models can be introduced by considering, at each split, only a
randomly drawn subset of all possible features. This gives rise to the random forest regressor
(RFR)~\cite{Breiman:ML45:5}. Thus, we see that RFR employs a 2-fold randomization procedure. The
ensemble can be made even more diverse by introducing further randomization in the learning process,
\eg, extremely randomized trees~\cite{Geurts:ML63:3}. RFR is an inherent ensemble method
encompassing diverse learners. This makes the model a very attractive option for a QBC-based
algorithm like \autoref{alg:QBC} or \autoref{alg:SQ}.

Another advantage of using the trees of an RFR in \autoref{alg:QBC} and \autoref{alg:SQ} is its
relatively low training complexity. An AL paradigm is a dynamic paradigm that needs to be performed
iteratively until we are satisfied with the performance. One wishes to be able to perform these
iterations quickly. Otherwise, the time saved from performing redundant electronic structure
calculations would be wasted in performing AL iterations. Building an RFR is relatively cheap. It
has an average time complexity of
$\Theta(M\cdot{}K\cdot\tilde{N}\log_2^2\tilde{N})$~[\onlinecite{Louppe:thesis:2014}] where $K,M$
denote the number of random features sampled at each splitting and the number of trees,
respectively, and $\tilde{N}\approx0.632N$ with the number $N$ of training examples.\footnote{The
   probability of not selecting a point in $n$ draws of $n$ samples with replacement is $(1-1/n)^n$,
   which converges in the limit of $n\to\infty$ to $e^{-1}$. Hence, bootstrap samples draw, on
   average, $1-1/e\approx63.2~\%$ of unique samples~[\onlinecite{Louppe:thesis:2014,
      Witten:DataMining:2017}]} This should be compared to the computational cost of training a GP,
which scales as $\mathcal{O}(N^3)$~\cite{Rasmussen:GaussProc:2006}. With extremely randomized trees
the average time complexity for training is
$\Theta (M \cdot K \cdot N \log_2 {N})$~\cite{Geurts:ML63:3}. The average inference complexity of
RFR is $\Theta (M \log {N})$~\cite{Geurts:ML63:3}. Thus, one AL iteration scales as
$\mathcal{O} (M * K*\tilde{N} \log_2 ^2\tilde{N})$ with the number of so-far-labeled data $N$. The
complexity of RFR is asymptotically inferior to that of a NN, which has a training time
complexity\footnote{This bound can be straightforwardly obtained by noting that matrix
   multiplications are the most expensive computations in the forward and backward passes of the NN
   training. We assume here that matrix multiplication scales as $\Theta (N^{3})$} of
$\mathcal{O}(N_e N(\sum_i^{l-1} N^{i} N^{i+1}))$ with the number of epochs $N_e$ needed for the NN
to converge and the number of neurons $N^{i}$ in layer $i$. However, in the data size regime of our
application, the computational costs of an RFR are smaller than
that of the NN.

\section{Simulations}
\label{sec:Simulations}
\begin{figure}
   \centering%
   \includegraphics[width=0.8\linewidth]{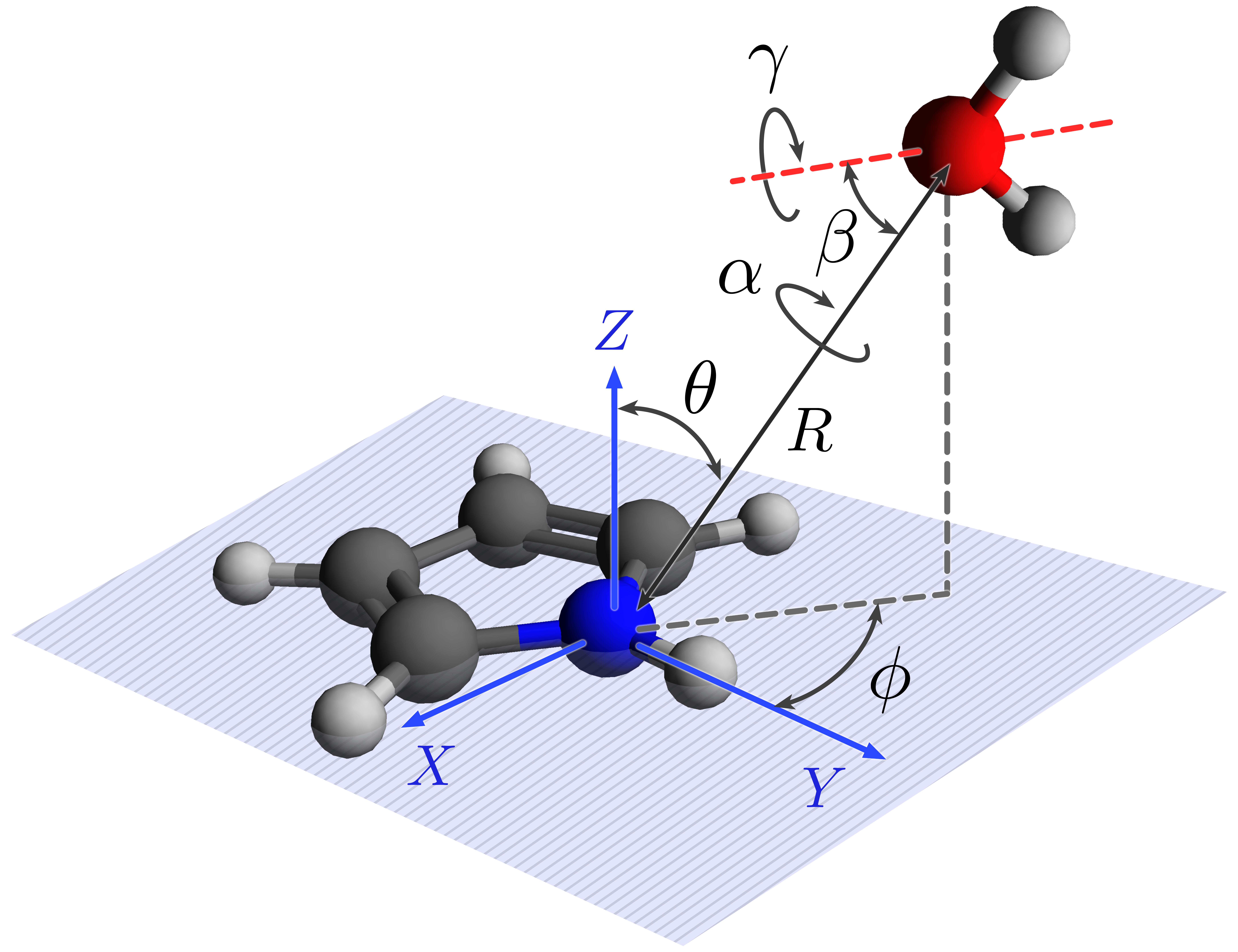}%
   \caption{Internal intermolecular coordinates $R$, $\theta$, $\phi$, $\alpha$, $\beta$, $\gamma$
      of \pyrrolew.}
   \label{fig:molecule}
\end{figure}
We applied different policy algorithms and tested their performance for building the prototypical
intermolecular PES of the \pyrrolew complex. Due to the highly fluxional nature of the hydrogen
bond, the intermolecular motions in \pyrrolew are highly delocalized, rendering the calculation and
representation of the PES very challenging. The intramolecular vibrations in the pyrrole and water
moieties can be described with a relatively simple, though multi-dimensional, single-minimum PES and
thus, for simplicity of calculations, were not considered here. We fixed the structures of pyrrole
and water monomers to the experimentally determined values~\cite{Tubergen:JPC97:7451,
   Nygaard:JMolStruct3:491}, see supplementary material, and
   varied the six intermolecular
coordinates, shown in \autoref{fig:molecule}. These are defined as follows: the relative position of
water with respect to pyrrole is described by the three spherical coordinates $R=[0.2,1]$~nm,
$\theta=[0,\pi]$, $\phi=[0,\pi]$ and the relative orientation of water is defined by the three Euler
angles $\alpha=[0,\pi]$, $\beta=[0,\pi]$, $\gamma=[0,\pi]$. The angles $\phi$, $\alpha$, and
$\gamma$ were restricted to the ranges $[0,\pi]$ exploiting the $\text{C}_\text{2v}$(M) symmetry of
the complex.

\begin{figure}
   \includegraphics[width=\linewidth]{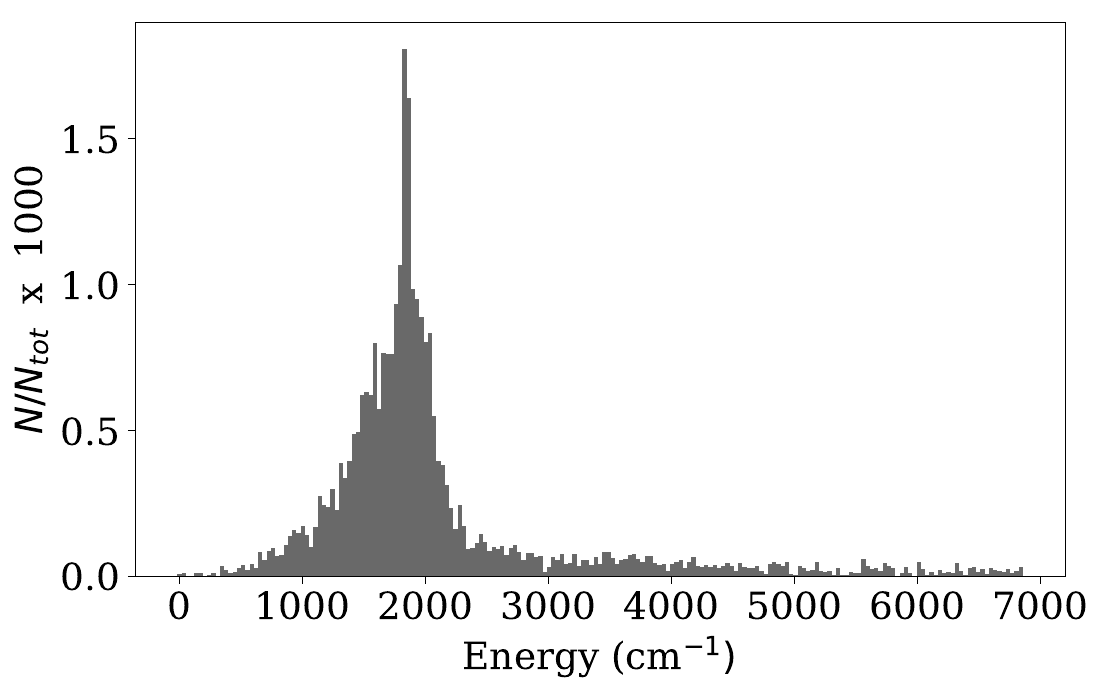}%
   \caption{The probability density distribution of the energies corresponding to all molecular
      geometries in the pool. The histogram was calculated for a bin width of $34.5~\invcm$ and has
      a peak at 1600~\invcm, corresponding to the dissociation limit of \pyrrolew.}
   \label{fig:dist_pool}
\end{figure}

The pool of molecular configurations was generated \emph{a priori} as the direct product of
one-dimensional grids for every degree of freedom and contained 57500 different molecular geometries
covering the potential energy up to 5000~\invcm above dissociation. All coordinates were sampled
more densely in the vicinity of the equilibrium geometry. Also, the angular coordinates were sampled
more densely for small radial distances $R\leq500$~pm with a sparser grid for $500<R\leq1000$~pm.
This led to a nonuniform distribution of energies in the pool, shown in~\autoref{fig:dist_pool}.
Note that a direct-product grid is not essential for the accumulation of the pool of unlabeled
geometries and the test dataset. Here, it was used mainly because it allows the coverage of the
whole configuration space that is relevant for the subsequent quantum dynamics' simulations, and
hence prevents biases and holes in the pool and test data. While this method is not arbitrarily
extendable to systems with more degrees of freedom, other pool accumulation
methods~\cite{Zhang:PRM3:023804} could be used without modifications to the SQBF approach. The
electronic structure calculations employed the density-fitting explicitly-correlated DF-MP2-F12
level of theory~\cite{Werner:JCP124:054114, Manby:JCP124:094103, Werner:JCP126:164102} in the
frozen-core approximation using aug-cc-pVDZ-F12~\cite{Peterson:JCP128:084102} atomic orbital,
cc-pVDZ-F12+/OPTRI~\cite{Shaw:JCTC13:1691} resolution of the identity, and
aug-cc-pVDZ/MP2FIT~\cite{Weigend:JCP116:3175} density fitting basis sets. The geminal exponent was
fixed at 1.0. The electronic structure calculations were carried out using
Molpro~\cite{Werner:JCP152:144107, Werner:WCMS2:242, MOLPRO}. A subset of 10~\% of the total number
of points in the pool was randomly selected as a test set and taken out of the pool (OOP). 5~\% of
the remaining data was randomly selected as a validation set. We employed
two different machine
learning models, RFR, and NN, to fit the data. We used exponential
functions of interatomic distances, with all distances
considered, as molecular descriptors, see \appref{ssec:tech}.

Furthermore, based on reviewer suggestions we tested the SQBF algorithm and an NN model on the PES
of the N$_4$ molecule using previously reported electronic structure
data~\cite{Paukku:JCP139:044309}.

\subsection{Performance}
\label{ssec:Performance}
For \pyrrolew we compared the performance of the RS, QBC, and SQBF AL policy algorithms considering
the convergence rate and the fitting accuracy. For QBC (\autoref{alg:QBC}) and SQBF
(\autoref{alg:SQ}) we used the trees of an RFR as an ensemble of learners. All policy algorithms
started from the same fixed amount of $m=2458$ labeled samples and queried the same equal number of
$m$ samples at every AL iteration. For every iteration and query algorithm, we used the RFR and NN
models to fit the data. The fitting error is defined as the RMS error of the ML models in predicting
the energies on the OOP dataset. This dataset was the same for all policy algorithms and followed
the joint distribution of the problem $\mathcal{P}$. The accuracy of a model on this dataset is an
estimate of the generalization error. Further technical details on the training process of the RFR
and NN models and their hyper parameters are provided in \appref{ssec:tech}.

\begin{figure}
   \includegraphics[width=\linewidth]{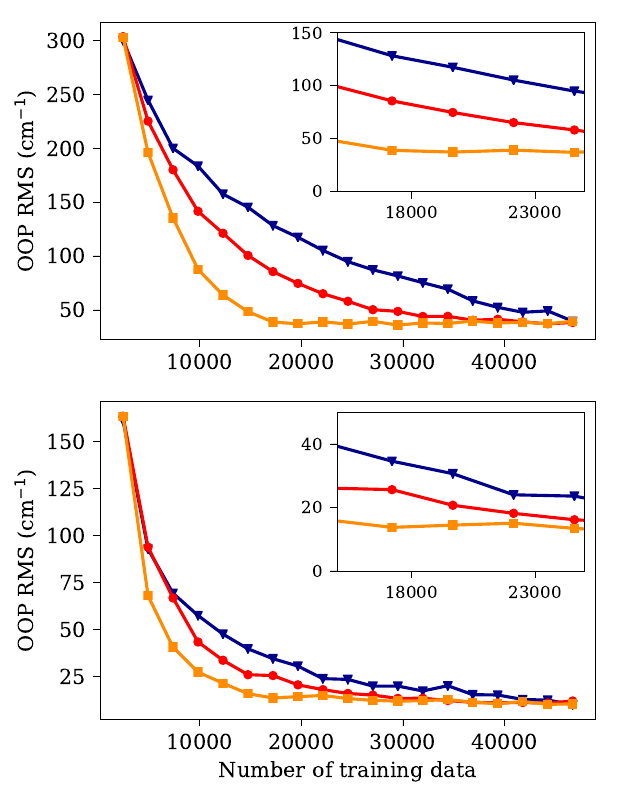}%
   \caption{RMS error of out-of-the-pool datasets using (up) the random forest regressor and (down)
      the neural network models for the RS (triangles), QBC (circles), and SQBF (squares) policies.
      The SQBF has the fastest convergence. A neural network model, trained on 30~\% of the total
      amount of datapoints in the pool achieves an RMS error of 16~\invcm. The RMS error on the full
      dataset is 11~\invcm. The neural networks trained on data collected by the QBC or RS
      algorithms show worse performance. The same convergence patterns hold when using a random
      forest regressor to train on the data instead of a neural network, albeit at overall somewhat
      slower convergence.}
   \label{fig:errorVSiter}
\end{figure}

The fitting errors of the RFR and NN models for different policy algorithms are plotted in
\autoref{fig:errorVSiter} as functions of AL iteration. The SQBF strategy with RFR model leads to
the fastest convergence of the error. QBC strategy outperforms RS. Similar convergence behavior of
different policy algorithms can be observed for the NN model. For our dataset, the fitting error of
NN was smaller than that of RFR for all AL iterations and for all strategies by an average factor of
3.3. \autoref{tab:RMSE} summarizes these results. The better performance of NNs is partially due to
the fact that NNs are easier to train to higher accuracy and can approximate complex functions with
a better control on the bias-variance trade-off, which was enabled by using an early stopping
criterion on the validation set, see \appref{ssec:tech}. The AL iterations were terminated when the
pool became empty. In practice, the iterations are to be terminated when the derivative of the
fitting error with respect to the amount of labeled data is less than a predefined
value~\cite{Zhai:JCP152:144103} or simply when the fitting error of the model is small enough.
\begin{table}[t]
   \caption{Out-of-the-pool RMS errors (in \invcm) of the random forest regressor and neural network
      models, listed as RFR/NN, computed for various fractions of the total pool data collected by
      the different AL policies}
   \begin{tabular}{lccccc}
     \hline\hline
     AL policy &20~\%&40~\% &60~\%&80~\%&100~\%		\\
     \hline
     RS   & 183/57 & 117/31 & 81/20 & 52/15 & 39/11 \\
     QBC  & 141/43 & 74/21  & 49/13 & 41/11 & 38/11 \\
     SQBF & 88/27  & 37/14  & 36/12 & 38/11 & 39/11 \\
     \hline\hline
   \end{tabular}
   \label{tab:RMSE}
\end{table}

Similarly, for the N$_4$ molecule the SQBF algorithm was used to query geometries from the pool of
$16421$ molecular geometries reported~\cite{Paukku:JCP139:044309}. The OOP and validation datasets
were each generated using $10~\%$ of the uniform-randomly sampled pool
data. An initial batch of $240$
geometries was uniform-randomly sampled from the pool and the SQBF algorithm queried 240 geometries
at each AL iteration. The same molecular descriptor as described above was used to transform the
data and an NN model was used for fitting; details on the NN design are provided in
\autoref{ssec:tech}. This procedure was repeated 100 times and the mean and standard deviation of
the resulting NN errors on the entire dataset as a function of the number of training examples is
reported in \autoref{tab:RMSE_N4}. The SQBF results are compared with the ensemble of 100 GPs used
to fit the data collected by the latin hypercube sampling algorithm~\cite{Cui:JPB49:224001}. The GP
method shows a better performance for the first few AL iterations. We attribute this to the fact
that it is hard to prevent overfitting with a neural network with a very small set of randomly
selected training data. However, already at 1200 training points the two models result in comparable
accuracy. With 1680 training points, our SQBF/NN approach achieves the same accuracy as GP with 2400
points, which corresponds to a $30~\%$ reduction in the size of the training dataset.
\begin{table}[t]
   \caption{RMS mean errors and standard deviations using the available data (in \invcm) of NN using
      the AL SQBF algorithm with a NN to fit the data (present work) and latin hypercube sampling
      with GPs to fit the data [\onlinecite{Cui:JPB49:224001}], applied to PES data of the N$_4$
      molecule [\onlinecite{Paukku:JCP139:044309}].}
   \begin{tabular}{cccccc}
     \hline\hline
     No. training data &NN (\invcm) & GP (\invcm) \\
     \hline
     240   & 36518 $\pm$ 2697 & 13300 $\pm$ 2770 \\
     480  & 26207 $\pm$ 1871  & 10027 $\pm$ 1371 \\
     720 & 11192 $\pm$ 1200  & 8401 $\pm$ 1102 \\
     960   & 8111 $\pm$ 668 & 7544 $\pm$ 972 \\
     1200   & 6201 $\pm$ 462 & 6806 $\pm$ 962 \\
     1680   & 4704 $\pm$ 612 & ---- $\pm$ -.-- \\
     1800   & 4494 $\pm$ 633 & 5551 $\pm$ 951 \\
     1920   & 4284 $\pm$ 658 & ---- $\pm$ -.-- \\
     2400   & 3557 $\pm$ 675 & 5012 $\pm$ 832 \\
     \hline\hline
   \end{tabular}
   \label{tab:RMSE_N4}
\end{table}
All the following further investigations are performed for \pyrrolew.
\begin{figure*}
   \includegraphics[width=\linewidth]{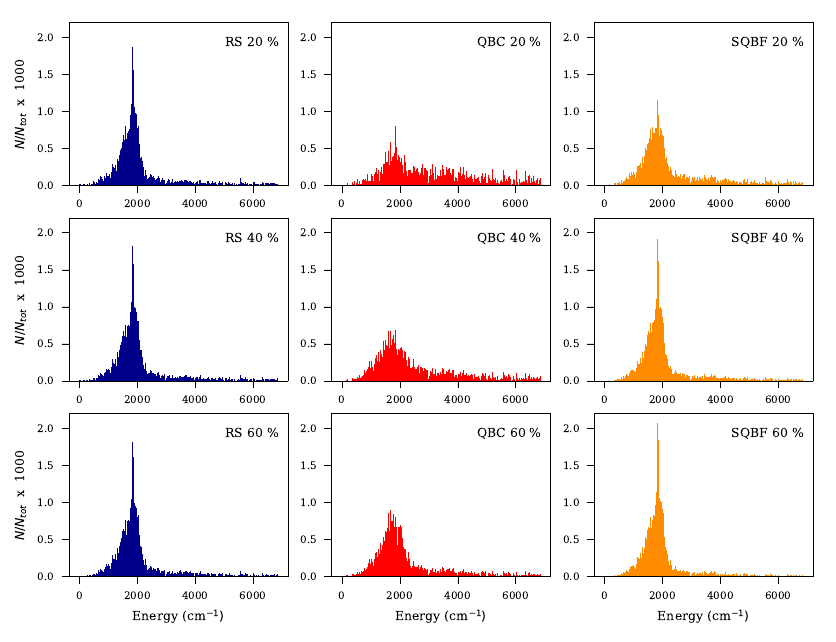}%
   \caption{Normalized probability density distributions of the number of data points
      $N/N_\text{tot}$ across the potential energies plotted for the data collected by the RS, QBC,
      and SQBF policies at different AL iterations corresponding to 20~\%, 40~\%, and 60~\% of the
      total pool. The bin width of the histograms is $34.5~\invcm$.}
   \label{fig:density}
\end{figure*}

\subsection{Distribution of queried data}
\label{ssec:view}
In \autoref{fig:density} we plotted the normalized distributions of the
samples' electronic energies of \pyrrolew
collected by different AL policies at three different iterations corresponding to $20~\%$, $40~\%$,
and $60~\%$ of the total pool. We compare these with the distribution of energies in the total pool
\autoref{fig:dist_pool}, which has a peak around 1600~\invcm, corresponding to the dissociation
limit of \pyrrolew. The densities were computed using 200 equally-sized bins covering the energy
range from 0 to 6874~\invcm and normalized to the bin width of $34.5~\invcm$. Evidently the
probability density of data sampled by the RS policy most closely resembles the pool distribution.
On the other hand, it is clear that the QBC algorithm samples more data with higher energies,
whereas SQBF keeps a balance between both the RS and QBC tendencies. As the number of the
labeled
data increases, all probability density distributions become more similar to the distribution in the
pool.

\begin{figure}
   \includegraphics[width=\linewidth]{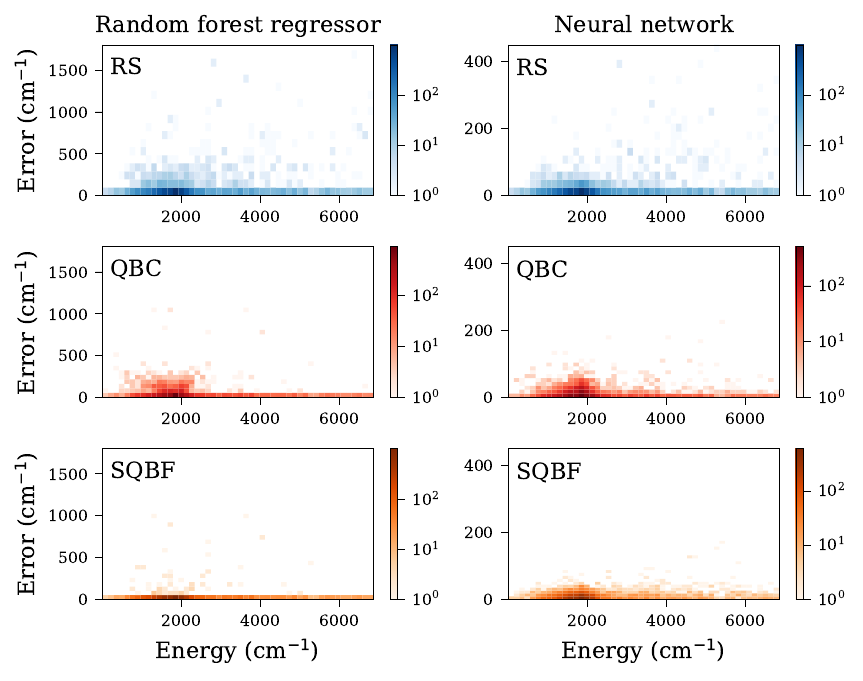}%
   \caption{2D histograms of discrepancies between the predictions of random forest regressor and
      neural network models (trained on 40~\% of the pool) and the potential energy of the
      out-of-the-pool data for different policy algorithms; 20 and 50 bins were used for energy and
      absolute error, respectively. Models trained on data collected by QBC tend to perform better
      on high-energy regions than on low-energy regions. The opposite is true for RS. In contrast,
      models trained on data collected by SQBF have a more uniform accuracy across the whole energy
      spectrum.}
   \label{fig:discrepancies}
\end{figure}

It is reasonable to expect that a model built on a dataset sampled by QBC algorithm will tend to
have a better performance for the high-energy regions. This is demonstrated
in~\autoref{fig:discrepancies} showing the 2D histograms of OOP energies and the absolute errors of
the RFR and NN models in predicting these energies, plotted for different policy algorithms. The
histograms were computed using 20 and 50 equally-sized bins for the energy and absolute errors,
respectively. The size of the training dataset here corresponds to $40~\%$ of the pool's size. We
clearly see that RS achieves good accuracy for the points with low energies, QBC works best for the
points with high energies, and the SQBF maintains a more regular accuracy across the whole energy
spectrum.

\subsection{Dependency on batch size and size of initially labeled dataset}
\label{ssec:Batch_Size}
\begin{figure}
   \includegraphics[width=\linewidth]{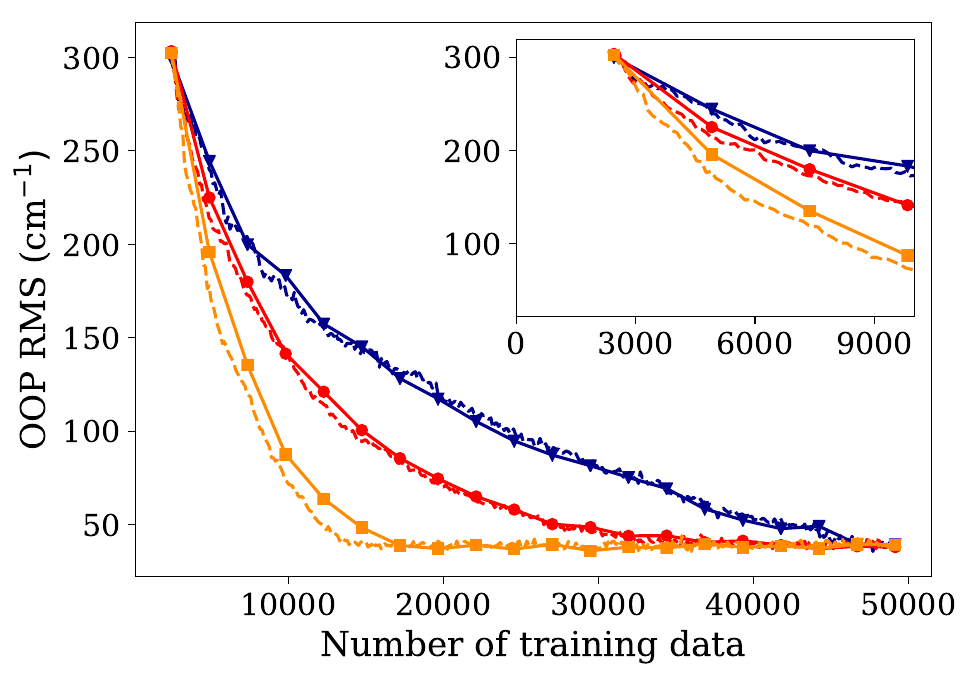}
   \caption{Effect of the batch size on the out-of-the-pool error of an RFR model trained using data
      collected by the RS (blue, triangles), QBC (red, circles), and SQBF (orange, squares)
      policies. Solid (points) and dashed lines correspond to the batch sizes fixed to 2458 and 122,
      respectively.}
   \label{fig:100iters}
\end{figure}

We repeated the above calculations with a smaller batch size of $122$ points instead of initially
used $2458$, starting from the same initially labeled dataset. The convergence of the RFR fitting
error with the number of training data are plotted in \autoref{fig:100iters} for different policy
algorithms. Here, we note that both QBC and SQBF strategies benefit slightly from using a smaller
batch size. This is in accordance with previous studies that showed a decreasing performance of QBC
with increasing batch size, which is due to collecting many similar samples~\cite{Ash:ICLR2020}.

\begin{figure}[!]
   \includegraphics[width=\linewidth]{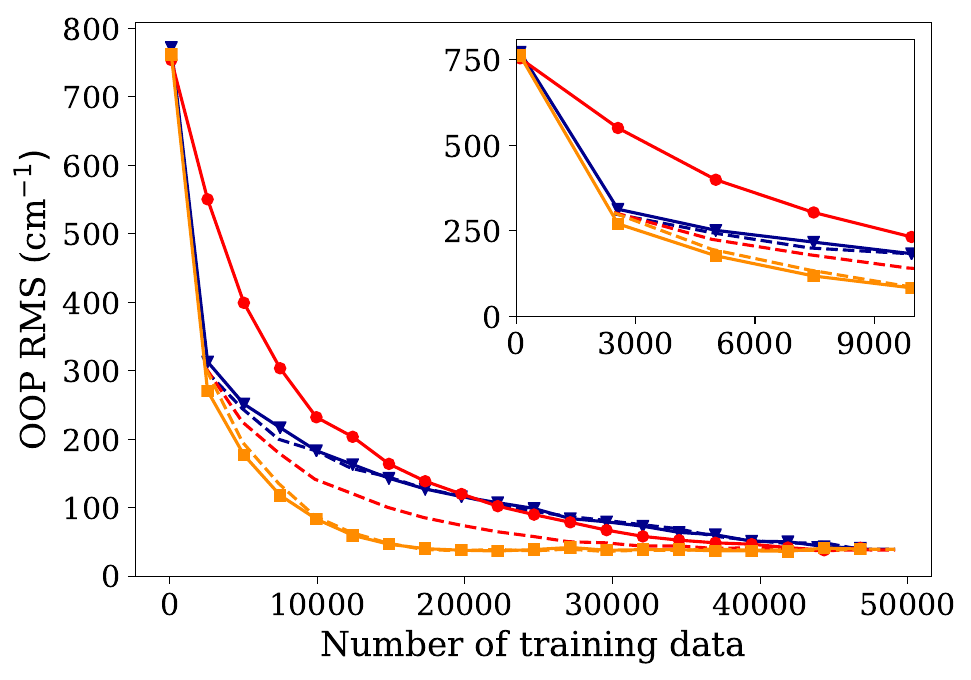}
   \caption{Effect of the size of initially labeled data on the the out-of-the-pool error of an RFR
      model trained using data collected by the RS (blue, triangles), QBC (red, circles), and SQBF
      (orange, squares) policies. Solid (points) and dashed lines correspond to 100 and 2458
      initially labeled data, respectively.} \label{fig:InitLabDat}
\end{figure}

We also studied the effect of changing the size of initially labeled dataset.
\autoref{fig:InitLabDat} shows the RFR fitting errors for different policy algorithms obtained from
initial datasets of $100$ and $2458$ samples with the batch size of $2458$. We observe that RS
policy
algorithm outperforms QBC, and that the accuracy of QBC declines significantly. This suggests that
with a fewer number of initially labeled data, an AL strategy should focus on collecting grid points
from dense regions of the configuration space rather than sampling points with high uncertainties in
their predictions. Notably, the SQBF performance is not affected by the size change.

\section{Summary and Conclusion}
\label{ssec:conclusions}
The first principles calculations of molecular PESs, especially for molecules with many fluxional
degrees of freedom, are computationally expensive. One of the major bottlenecks originates from the
need to perform the expensive quantum chemical calculations for tens and hundreds of thousands of
different molecular geometries. Algorithms that allow to reduce the number of necessary single-point
calculations with controlled accuracy of the resulting PES are thus highly demanded. For small
molecules, grid reduction algorithms were found beneficial in calculations employing high-level
electron correlation, basis set, and relativistic corrections, which are usually computationally
affordable only for a relatively small number of points~\cite{Peterson:TCA131:1079,
   Owens:JCP142:244306, Yachmenev:JCP135:074302, Dral:JCP152:204110}.

We developed an accurate and automated procedure to efficiently sample the molecular geometry grid
points leading to a systematic convergence of the accuracy of PES. We employed a regression version
of SQBF, a pool-based active learning algorithm to generate a compact grid of molecular geometries
and the RFR and NN ML-models to construct the six-dimensional intermolecular PES of the weakly-bound
\pyrrolew complex. The proposed method samples grid points with high uncertainties in the
corresponding predictions of energies and at the same time preserves the statistical information
embedded in the pool. In our benchmark application to the six-dimensional intermolecular PES of
\pyrrolew this led to a roughly two times faster convergence with respect to the number of grid
points than the commonly used QBC algorithm to represent the PES to an accuracy of about 16~\invcm.
Furthermore, the PES fitted on the data sampled by SQBF exhibited a more uniform accuracy across the
whole energy spectrum in comparison to QBC. We empirically showed that the SQBF method is not very
sensitive to a variation of parameters such as the size of initially labeled data and size of the
batch.

In addition, the proposed method is computationally cheap and scales well with the size of the
labeled data $N$, \ie, as $\Theta(M\cdot{}K\cdot{}\tilde{N}\log_2^2\tilde{N})$, where $K,M$
denote
the number of random features sampled at each splitting and the number of trees, respectively,
$\tilde{N}=0.632N$. This makes the method attractive for developing universal ML-potentials where
large datasets are needed~\cite{Ramakrishnan:SD1:1, Rupp:PRL108:058301, Smith:SD4:170193}. In the
case when the accuracy of the RFR is not sufficient for the application of the PES, we showed that
the data can be used equally-well by other ML models like NNs. An alternative would be to
employ~\autoref{alg:SQ} with any other ensemble of learners or even with a model that offers a
direct computation of uncertainty.

Overall, the presented procedure is general and can be applied to the PESs of any polyatomic
molecule. It can also be used to model other physical properties like dipole-moment or
polarizability surfaces. The major advantage of the proposed method over more popular QBC approach
is the heuristic sampling procedure that preserves the distribution of data in the pool while
keeping the uncertainty as the primal selection criterion. We believe that in the future the general
approach can be improved even further by a better tuned balance between uncertainty and
representativeness.
\section*{Supplementary Material}
See supplementary material for the computed potential energies of 
pyrrole(H2O), and the structures of pyrrole and water
monomers.
\section*{Acknowledgment}
This work has been supported by Deutsches Elektronen-Synchtrotron DESY,
a member of the Helmholtz Association (HGF), the Data Science in Hamburg HELMHOLTZ Graduate School
for the Structure of Matter (DASHH, HIDSS-0002) and by the Deutsche Forschungsgemeinschaft (DFG)
through the cluster of excellence ``Advanced Imaging of Matter'' (AIM, EXC 2056, ID 390715994). We
acknowledge the use of the Maxwell computational resources operated at Deutsches
Elektronen-Synchrotron DESY, Hamburg, Germany.

\section*{Data availability}
The data that supports the findings of this study are available within the article and its
supplementary material. The computer code and all relevant data for this work can be obtained from
the git repository at
\href{https://github.com/CFEL-CMI/Active-Learning-of-PES}{https://github.com/CFEL-CMI/Active-Learning-of-PES}.

\appendix
\section*{Appendix}
\subsection{Technical details on the random perturbations in AL algorithms and the training of
   machine learning models}
\label{ssec:tech}
The regression trees used to implement the QBC and SQBF algorithms were both built using the
scikit-learn (\texttt{sklearn}) Python package~\cite{Pedregosa:JMLR12:2825}. All AL algorithms used
here were written based on the Libact Python package~\cite{Yang:arXiv:1710.00379}. For both the QBC
and SQBF algorithms we used an ensemble of 100 trees. The training during all AL iterations used an
exponential function of the inter nuclear distances as a molecular descriptor: {$1-\exp(-(r-r_0))$}
where $r,r_0$ denote the actual distance and equilibrium distance between two atoms, respectively.
The perturbation of the learning process is controlled through two parameters: (i) a bootstrapping
parameter $\gamma$ that determines the fraction of data sampled by each tree and (ii) the number of
features $\beta$ sampled randomly by each tree. For the batch sizes used in the simulations we
experimented with several combinations of these parameters and obtained the best convergence for
$\gamma=0, \beta=12$ for simulations on \pyrrolew and $\gamma=0,
\beta=4$ for simulations on N$_4$. The same parameters were used for
uncertainty estimation in both the QBC and
SQBF algorithms. Minimal cost complexity pruning was used to reduce the overfitting of RFR with
complexity parameter $c=0.01$. Since we require that \autoref{alg:SQ} queries exactly $B$
geometries, one may run into a situation where the number of entries with non-zero probabilities of
the distribution $p$ is less than $B$. In such a case we chose to query the elements with highest
uncertainty. The effect of this choice on the simulations conducted in the manuscript is negligible
since we only encountered this case once in one of the last AL iterations.

The NN used is a multilayer perceptron and training was implemented using the Python Tensorflow
package~\cite{Abadi:Tensorflow:2015}. The NN has three hidden layers with 256, 512, and 256 neurons,
respectively, and a single neuron output layer. The second and third layers were $l_2$-regularized
with a regularization parameter of $10^{-5}$. All hidden layers used ``ReLU'' as the activation
function. The ReLU activation function could be substituted with a smooth approximation such as
Softplus to yield a smooth PES. The same aforementioned molecular descriptor was used. The networks
were trained for 250 epochs using the Adam optimization algorithm~\cite{Kingma:ICLR3:2015}, with an
initial learning rate of $0.0025$ and a decaying learning rate schedule
($\text{lr}_{\text{current}}=0.9825\times\text{lr}_{\text{previous}}$). An early stopping callback
was employed on the validation set that was taken out-of-the-pool with patience of $25$. The NN
hyper parameters were set to obtain a sufficiently accurate NN, with test error of around 10~\invcm
when using all the training data.

\bibliography{string,cmi}%
\onecolumngrid%
\listofnotes%
\end{document}